\documentclass[runningheads]{llncs}
\usepackage{graphicx}
\usepackage[dvipsnames]{xcolor}
\usepackage{multirow}
\usepackage{tabularx}
\usepackage{wrapfig, booktabs}
\usepackage{amsfonts}
\usepackage{amsmath}
\usepackage{graphicx,verbatimbox}
\usepackage[breaklinks]{hyperref}

\begin{document}
\title{BeSt-LeS: Benchmarking Stroke Lesion Segmentation using Deep Supervision}
\titlerunning{BeSt-LeS: Benchmarking Stroke Lesion Segmentation}

\author{Prantik Deb\inst{1} \and
Lalith Bharadwaj Baru\inst{1} \and  Kamalaker Dadi\inst{1} \and Bapi Raju S\inst{1}}

\authorrunning{Prantik Deb \emph{et al.}}
\institute{
$^1$Brain Cognitive Computation Lab, Cognitive Science\\ IIIT, Hyderabad-32\\
\email{prantik.deb@ihub-data.iiit.ac.in}\\
\email{lalith.baru@research.iiit.ac.in}\\
}



%

\maketitle              
\begin{abstract}
Brain stroke has become a significant burden on global health and thus we need remedies and prevention strategies to overcome this challenge. For this, the immediate identification of stroke and risk stratification is the primary task for clinicians. To aid expert clinicians, automated segmentation models are crucial. In this work, we consider the publicly available dataset ATLAS $v2.0$ to benchmark various end-to-end supervised U-Net style models. Specifically, we have benchmarked models on both 2D and 3D brain images and evaluated them using standard metrics. We have achieved the highest Dice score of 0.583 on the 2D transformer-based model and 0.504 on the 3D residual U-Net respectively.  We have conducted the Wilcoxon test for 3D models to correlate the relationship between predicted and actual stroke volume. For reproducibility, the code and model weights are made publicly available. 
\end{abstract}

\keywords{Stroke Lesion Segmentation  \and T1 Weighted MRI \and Deep Supervision \and ATLAS $v2.0$ \and Deep Learning}
\section{Introduction}
Brain stroke has become a significant burden on global health with increasing prevalence in low- and middle-income countries. Therefore there is an urgent need for targeted prevention strategies and improved healthcare infrastructure to address this growing public health challenge \cite{johnson2016stroke},  \cite{kim2020global}. Given brain Magnetic Resonance (MR) images on stroke populations, localizing and detecting the lesions is crucial for clinicians. However, the automation of the localization process has achieved significant reach \cite{baird1998magnetic} with novel machine-learning models that can aid clinicians. To fuel these models, we need datasets that could automatically segment to the level of expertise and doing so, it could ease the clinician's task.

In this context, ATLAS (Anatomical Tracings of
Lesions After Stroke) v1.2 dataset made progress by creating  304 T1-weighted MRI samples collected from 11 cohorts. This ATLAS v1.2 \cite{liew2018large} was released in 2018 and of which, 229 standardized subjects were available with T1-weighted MRI image and its corresponding lesion mask. Later in 2022, ATLAS v2.0 \cite{liew2022large} was released and it has 1217 T1-weighted MRI samples collected from 44 cohorts of which 655 samples mask were availed to the public. In specific, there was an extra margin of 426 samples from version 1.2 to 2.0.  

There were numerous impressive models that performed well on ATLAS v1.2. Whereas, ATLAS v2.0 was released recently and therefore, there is quite less progress. Therefore, we outpace and benchmark ATLAS v2.0 on various standard U-Net style architectures for both 2D and 3D brain images. 

In Table \ref{tab:novelty2d_3d}, ATLAS v1.2 was applied using distinct U-Net architectures for 2D modality but, there isn't any model to date that has been implemented on ATLAS v2.0. Similarly, in the case of 3D modality, Table \ref{tab:novelty2d_3d} illustrates there are few implementations for ATLAS v2.0 using nnU-Net as their underlying framework \cite{isensee2021nnu}. Thus, we contribute by analyzing the ATLAS v2.0 dataset for both 2D and 3D modality. A brief description of each of the models is detailed in the supplementary material.

\begin{table}[!t]
\caption{The table summarizes the glimpse of 2D and 3D U-Net variants whether they were implemented on the ATLAS v1.2 and v2.0 T1 MR images denoted as Yes or No. The implementation of 2D models on ATLAS v2.0 is sparse.}
\begin{center}
\begin{tabular}{|c|c|c|}
\hline
\bf 2D Models & \bf ATLAS v1.2 & \bf ATLAS v2.0  \\ \hline \hline
U-Net & Yes \cite{yu2023san} \cite{qi2019x} \cite{zhou2019d}\cite{zhang2020mi} \cite{sheng2022cross} & No\\
Residual U-Net & Yes \cite{yu2023san} \cite{qi2019x} \cite{sheng2022cross} & No\\
         Attention U-Net & Yes \cite{sheng2022cross} & No\\
         Transformer Based & Yes \cite{wu2023w} & No \\ \hline
          \bf 3D Models &\bf ATLAS v1.2 &\bf ATLAS v2.0  \\ \hline \hline
         U-Net & Yes \cite{zhou2019d} \cite{zhang2020mi}  \cite{paing2021automated} & Yes \cite{verma2022automatic} \cite{huo2022mapping}\\
         Residual U-Net & Yes \cite{tomita2020automatic} & Yes \cite{huo2022mapping}\\
         Attention U-Net & No & No\\
         Transformer Based & No  & No \\ \hline
\end{tabular}\label{tab:novelty2d_3d}
\end{center}
\end{table}

\section{Contributions of this work}
\begin{enumerate}
    \item To the best of our knowledge, ours is the first attempt to benchmark the standard segmentation models \emph{i.e.} both convolution and transformers-based architectures on the ATLAS v2.0 dataset. 

    \item We have also conducted experiments for both 2D and 3D-based models. We report our highest dice score of 0.58 on the 2D transformer-based model. Also, we have achieved a 0.504 dice score on the 3D residual U-Net. 

    \item Finally, we conduct the Wilcoxon test on 3D models and compare the relationship between predicted and actual stroke volume. 
\end{enumerate}
\section{Data and Models}
\label{Methods}
This section briefly discusses the dataset and methods considered for analysis. The organization is as follows: First, we introduce the dataset and the training strategy implied. Next, we detail the significance of various U-Net style models.
\subsection{Dataset}
\begin{wraptable}{r}{6.0cm}
\caption{The below table describes the train validation and test proportions divided for training supervised 2D U-Net style architectures. The number of samples below represents the number of subjects. The slices are unevenly divided based on the volume of the T1-weighted MRI and the axial plane is considered while cropping each slice.}\label{tab-data2d}
\centering
\scalebox{0.99}{
\begin{tabular}{cccc}\\
\toprule 
Split & \% & Samples & Slices  \\
\hline\hline 

Train      & 60 & 393 & 15394 \\  \hline
Validation & 20 & 131  & 4666 \\  \hline
Test       & 20 & 131  & 5452 \\  \bottomrule
\end{tabular}
}

\caption{In the below tabular data, the train validation and test proportions are divided for training supervised 3D U-Net style architectures. The number of samples below represents the number of subjects.}\label{tab-data3d}
\centering
\scalebox{0.99}{
\begin{tabular}{ccc}\\\toprule  
Split & \% & Samples \\\hline\hline

Train      & 60 & 393\\  \hline
Validation & 20 & 131\\  \hline
Test       & 20 & 131\\  \bottomrule
\end{tabular}
}
\end{wraptable} 
The ATLAS (Anatomical Tracings of Lesions After Stroke) data consists of T1-weighted MRI images of subjects having lesions due to stroke. This data has two versions, ATLAS v1.2 \cite{liew2018large}, and ATLAS v2.0 \cite{liew2022large}, respectively. For our analysis, we solely conduct our experiments on ATLAS v2.0 dataset \cite{liew2022large}, which is publicly available. The samples in this data (ATLAS v2.0) were collected from 44 diverse cohorts with a total sample size of 1271. From these 1271 samples, only 655 samples consist of image-to-mask pairs dedicated to training the models. Another 300 samples are treated as hidden-set and do not reveal the masks of T1-weighted MRI images\footnote{Additional information such as lesion numbers, cortical location, and severity of stroke for each subject can be found in the original paper}. 

While analyzing ATLAS v2.0 we conduct experiments both on 2D and 3D modality. As masks for the original test set are inaccessible we divide the training samples (655) into train validation and test proportions. For 3D modality, the data can be directly fed to the model. But for 2D modality, each subject's 3D T1-weighted MRI images are to be cropped into slices along the z-axis (axial) and then given as input to 2D U-Net style architectures. For both the modalities the \emph{Z}-score Normalization are performed as a pre-processing step \cite{akkus2017deep}. The information regarding train, test, and validation sets are elucidated in Table \ref{tab-data2d} and \ref{tab-data3d}. For the 2D dataset, the discrepancy between the validation and the test is due to a rejection of 0.1\% lesions in the given 2D slice. Specifically, wherever the slices and their respective mask pairs were rejected were not included in the study. Now, this data is to be processed using various segmentation architectures to segment the lesion in the T1-weighted MRI images in the protocol mentioned in Table \ref{tab-data2d} and \ref{tab-data3d}. 

\subsection{U-Net Style Architectures}
The delineation of a specific organ or certain tissue site or cell nuclei from given medical images is one of the crucial tasks in medical image analysis. Various deterministic algorithms were developed to automate this process, some of which are random walks \cite{grady2006random} and SLIC \cite{achanta2012slic}. But later with the aid of deep learning, more sophisticated and \emph{learnable} methods were developed \cite{ciresan2012deep}. Later, Ronneberger \emph{et al.} \cite{ronneberger2015u} proposed U-Net architecture from which the field of medical image segmentation caught its attention as U-Net was fast, modulable, and robust.  Later there were many methods that were crafted using the U-Net style as the underlying framework. Thus, we study and benchmark some of the fundamental U-Net style architectures that achieved significant results in the field of medical image segmentation.

\begin{figure}[!t]
\includegraphics[width=\textwidth]{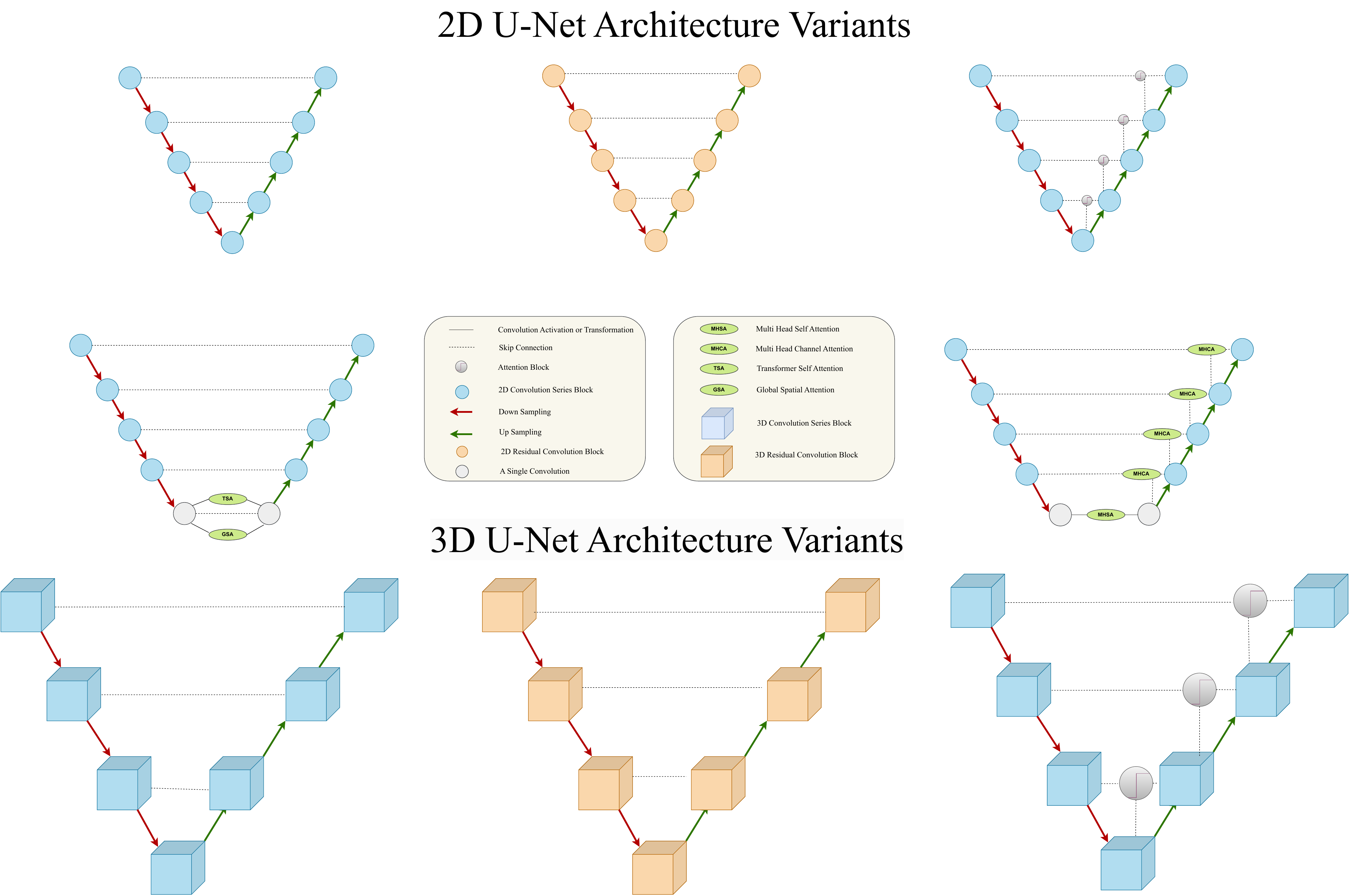}
\caption{The figure illustrates various U-Net style architectures. (First row) shows a diagrammatic view of the convolution-based Transformer models and (bottom row) shows two novel transformer-based U-Net architectures. We detailed all the symbols and signs used in the legend block.} \label{unets}
\end{figure}

\paragraph{\bf U-Net:} This was the first deep learning architecture that superseded the existing models with large margins both computationally and performance-wise. This architecture, thus, was made as a baseline for many medical segmentation tasks \cite{ronneberger2015u}. Soon after the U-Net, the architecture was modified to learn volumetric data using 3D convolutions \cite{cciccek20163d} with sparse annotations. The architecture is very similar to that of 2D U-Net except that, 3D can process volumetric information using 3D convolutions and it is depicted in the last row of Fig. \ref{unets}.

The architecture is quite intuitive, as features are downsampled using pooling layers to a latent space or base and later upsampled to reconstruct the desired mask image. This latent space is perceived to preserve the crucial features, while the skip-connections (these are denoted as ($\cdots$)) aid the reconstruction by guiding to map of the structural information. In the Fig. \ref{unets}, you can see each \textcolor{SkyBlue}{blue} circle represents a series of convolution layers, and the red arrow (\textcolor{red}{$\rightarrow$}) indicates that an image of size $h \times w$ is downsampled (pooled) to $\frac{h}{2} \times \frac{w}{2}$. The downsampling operation is quadruply performed to get to the latent space. Now, from this latent space, the acquired features are upsampled quadruply (indicated with a green arrow (\textcolor{green}{$\rightarrow$}))  to produce the desired segmentation.

\paragraph{\bf Residual U-Net:}
He \emph{et al.}\cite{he2016deep} proved that residual connections tend to provide refined representations for downstream tasks with less computation and better performance. In this regard, Zhang \emph{et al.} \cite{zhang2018road} proposed a U-Net architecture with residual connections for extracting road patterns from aerial imagery. This later was implied in the domain of medical images by Alom \emph{et al.}\cite{alom2019recurrent} with additional memory components. Later, the residual connections were implied between a series of 3D convolutions to produce volumetric segmentation \cite{yu2019liver}, \cite{bhalerao2019brain},\cite{isensee2019attempt}. The architecture is very similar to that of 2D residual U-Net except that, 3D can produce volumetric masks from 3D medical data, and the pictorial interpretations are illustrated in the last row of Figure \ref{unets}.

The current Residual U-Net style was inspired by Zhang \emph{et al.}\cite{zhang2018road}. In U-Net, before downsampling at each step, there are a series of convolutions with residual connections, which are represented by an \textcolor{orange}{orange} circle. The rest, upsampling (\textcolor{green}{$\rightarrow$}) and downsampling (\textcolor{red}{$\rightarrow$}) operations, are similar to that of traditional U-Net.

\paragraph{\bf Attention U-Net:}
The fundamental concept of \emph{attention} was formulated by Bahdanau \emph{et al. }\cite{Bahdanau2014Neural}. Later Oktay \emph{et al.} \cite{oktay2018attention}  applied this mechanism as '\emph{Attention Gate}' (AG), which improved segmentation with detailed localization of multiple organs.

In this architecture, the core component is the attention gate which aids the U-Net in segmenting desired lesions. The architecture style in downsampling is similar to U-Net, i.e., the image is quadruply downsampled (\textcolor{red}{$\rightarrow$} $\times 4$) to obtain the latent space. In traditional U-Net, upsampling (\textcolor{green}{$\rightarrow$}) is achieved using transpose convolutions, and additional representations are aggregated from the skip connections. But in Attention U-Net, the representations from the previous layers and from skip connections are aggregated using AG, and now these features are concatenated with the upsampled (\textcolor{green}{$\rightarrow$}) features. Thus, they form cascaded connections resulting in better segmentation. Similarly,  this attention gate is applied to 3D convolutions to get volumetric attention \cite{nodirov2022attention}, \cite{wang2019volumetric}.


\begin{table*}[!t]
\centering
\caption{The below table illustrates the performance of variants of 2D U-Net architectures. The first three models are pure convolution-based architectures and the remaining two are hybrid networks with convolutions and transformer components. The evaluation criteria implied is the same as Table \ref{tab-data2d}; We report the performance of the model for the test set.}\label{res-2dunets}
\begin{tabular}{|c|c|c|c|c|}
\hline
\multirow{2}{*}{\textbf{Method}} & \multicolumn{4}{c|}{\textbf{Performance Metrics (2D Data)}} \\ \cline{2-5}
& Dice Score & IoU Score & Precision & Recall \\ \hline \hline
U-Net \cite{ronneberger2015u} & 0.417 & 0.337 & 0.580 & 0.360 \\
Residual U-Net \cite{zhang2018road} & 0.456 & 0.375 & 0.592 & 0.420 \\
Attention U-Net \cite{oktay2018attention} & 0.487 & 0.396 & 0.636 & 0.439 \\ \hline
TransAttn U-Net \cite{chen2021transattunet} & 0.572 & \textbf{0.477} & \textbf{0.660} & 0.565 \\
U-Net Transformer \cite{petit2021u} & \textbf{0.583} & 0.475 & 0.659 & \textbf{0.591} \\ \hline
\end{tabular}
\end{table*}

\paragraph{\bf TransAttn U-Net:}
This architecture was designed by Chen \emph{et al.} \cite{chen2021transattunet} in which they propose SAA: Self-Aware Attention, which is an amalgamation of multi-level and multi-scale guided attention mechanisms. In specific, after downsampling features to the embedding space, they perform two attention mechanisms which are Transformer Self-Attention ($\mathcal{F}_{TSA}$) and Global Spatial Attention ($\mathcal{F}_{GSA}$). Now, these features are combined into a single convolution block, and with each step of upsampling, the previous layer features are attached as skip connections using \emph{Bi-linear Upsampling} (refer Fig. \ref{unets}). The significance of each attention mechanism is elucidated below.  

Suppose our image is represented as $X \in \mathbb{R}^{t \times h \times w}$, where $t, h, w$ are time-steps (channels), height, and width of the given image, respectively. The image is passed to the encoder, and the downsampled representation is denoted by $\mathcal{F}_{base}^t \in \mathbb{R}^{t \times (h \times w)}$. Now, to achieve GSA, 

\begin{equation}
    \mathcal{F}_{GSA}(M,N,W)_i = \sum_{k=1}^{h\times w} \left( W_k \mathcal{A}_{i,j} \right)
\end{equation}

Where, $N \in \mathbb{R}^{t' \times (h \times w)}$ and $M \in \mathbb{R}^{ (h \times w) \times  t'}$. Also the $\mathcal{A}_{i,j} = \frac{e^{(M_i N_j)}}{\sum_{r=1}^n e^{(M_r N_j)}}$ . This $\mathcal{A}_{i,j}$ measures the input of the $i^\text{th}$ and $j^\text{th}$ position. Similarly, the TSA attention is calculated as,

\begin{equation}\label{kqv-eq}
    \mathcal{F}_{TSA}(K,Q,V) = soft\left(\frac{QK^T}{(d_k)^{1/2}} \right) V
\end{equation}

Where $K, Q, V$ are just the features of $\mathcal{F}_{base}^t$ added with positional encoding and $d_k$ is the dimensionality of any key or Query or value sequence (i.e., $d_k = |V| or |Q| or |K|$ and $soft(.)$ is the softmax activation function \cite{elfadel1993softmax}.) These attentions are operated at the latent space or base, and now, in final step, all these features are amalgamated at the latent space as,

\begin{equation}
    \mathcal{F}_{SAA} = \psi_1 \mathcal{F}_{TSA} + \psi_2 \mathcal{F}_{GSA} + \mathcal{F}_{base}
\end{equation}

Where $\psi_1$ and $\psi_2$ are the scale parameters,  respectively, and they control the importance assigned to each attention mechanism. Initially, they are assigned with null weights and gradually incremented to obtain a systematic consistency.

\paragraph{\bf U-Net Transformer:}
This method originated from the work by Petit \emph{et al.} \cite{petit2021u}. The authors impart self-attention and channel attention modules in this work to produce interpretative segmentation throughput. Fundamentally, the self-attention module uses Multi-head Self-Attention (MHSA) which is similar to Vaswani \emph{et al.} \cite{vaswani2017attention}, and this aims to acquire long-range structural information from the images that were downsampled to the latent-space. The underlying operation is quite similar to equation (\ref{kqv-eq}).

\begin{figure}[!t]
\includegraphics[width=\textwidth]{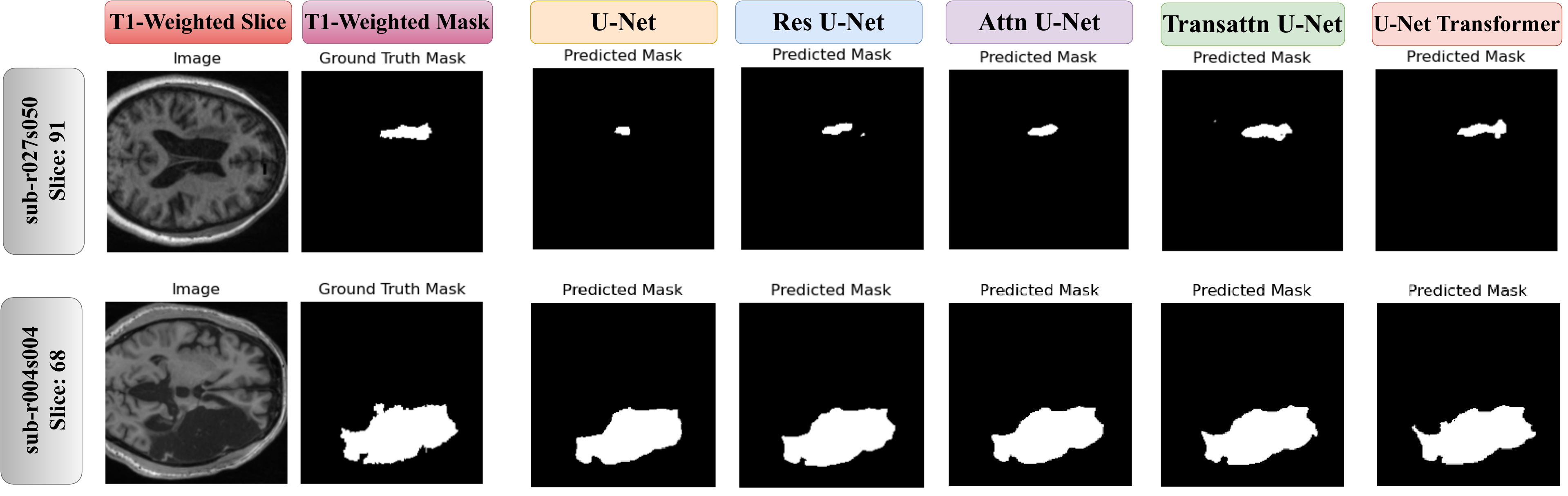}
\caption{2D visualizations of the benchmarks between ground truth and predicted lesions for two subjects. As can be seen, we have two different sizes of stroke lesion subjects included for visualization (Subject ID: $sub-r027s050$ slice 91 and $sub-r004s004$ slice 68) of which one is small and the other being large. We display the predicted outputs of convolution and transformer-based 2D U-Net models for these subjects. All 2D models performed equally better but the U-Net transformer gave the finest boundaries as visible for both the subjects.} \label{fig-2dunetplot}
\end{figure}

In the channel-attention module, the representations from skip connections (at each pooling step) are first applied with MHSA and then concatenate the features coming from latent space after each upsampling step. Thus this module is referred to as Multi-Head Channel-Attention (MHCA) as initially the input is transferred to MHSA and then concatenated with a cross-attention mechanism. The diagrammatic explanations are elucidated in Fig. \ref{unets}.


\begin{table*}[t!]
\centering
\caption{The table illustrates the performance of variants of 3D U-Net architectures. These models are pure convolution-based architectures, and the evaluation criteria are the same as in Table \ref{tab-data3d}. We report the performance of the model on the test set.}
\label{res-3dunets}
\begin{tabular}{|c|c|c|c|c|} 
\hline
\textbf{Method} & \multicolumn{4}{c|}{\textbf{Performance Metrics (3D Data)}} \\ \cline{2-5}
& Dice Score & IoU Score & Precision & Recall \\ \hline \hline
U-Net \cite{cciccek20163d} & 0.450 & 0.350 & 0.584 & 0.444 \\
Residual U-Net \cite{zhang2018road} & \textbf{0.504} & \textbf{0.393} & \textbf{0.585} & 0.533 \\
Attention U-Net \cite{oktay2018attention} & 0.469 & 0.369 & 0.498 & \textbf{0.578} \\ 
\hline
\end{tabular}
\end{table*}



\section{Results and Discussion}
In this section, we experiment with the aforementioned methods as summarized in Section \ref{Methods}. First, we evaluate the performance of 2D and 3D models. Later, we conducted the Wilcoxon test using the ground truth and predicted lesion volume for 3D U-Net models. 

\subsection{Results for 2D}
Among the 2D U-Net convolution-based models the attention U-Net has proven to have a significant Dice score of 0.487 (with an extra verge of 7.0\% dice score from baseline\footnote{In this paper we consider, standard U-Net to be our baseline for both 2D and 3D models respectively.}). Whereas the hybrid transformer- and convolution-based models tend to provide a noteworthy performance of 0.583 and 0.572 dice scores. These models superseded 2D standard U-Net with an additional 15.5 and 16.6\% of dice score respectively. All these results are illustrated in Table \ref{res-2dunets}.

As there were no additional augmentations applied we can directly infer that adding self-attention components (latent space) such as TSA, MHSA, and GSA played a crucial role in providing significant performance with their cascaded attention \cite{wu2023w}. 

In this regard, we have considered two subjects for visualizing the performance of each model in predicting lesions. In Figure \ref{fig-2dunetplot} we have displayed the potential of each model to segment small lesions (first row) and large lesions (second row). All the models were near good in segmenting the large volume of lesions but the U-Net Transformer is able to accurately delineate the boundaries of tissues from T1-weighted MRI. But, most of the models struggle to extract the small lesions. Hybrid models such as transformer-based architecture did perform well (comparatively) to a certain extent but, still, there is a wide scope for developing novel models. 



\begin{figure}
\includegraphics[width=\textwidth]{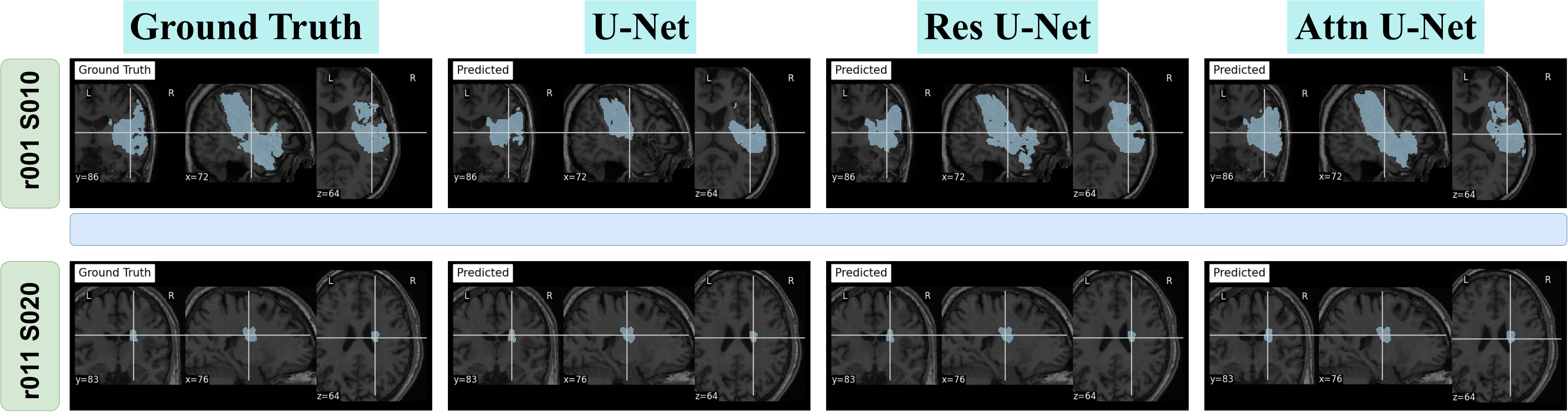}
\caption{The above visualization was considered from the test set (ID: $sub-r001s010$ and $sub-r011s020$ ) and compares three 3D U-Net models (standard, Residual, and Attention). For each model, the left part remains as ground truth and the right part is the model's predictions. In each image, the visualization elucidates the precise location of stroke in the brain using three axes (sagittal, coronal, and axial).} \label{fig-3dunetplot}
\end{figure}

\subsection{Results for 3D}
Now, we consider 3D U-Net style architectures which include standard U-Net, Residual, and Attention U-Net. Due to the added temporal relationship among the features in 3D convolution, the standard 3D U-Net was able to achieve a 0.45 dice score. Previously, in 2D modality, the convolution-based model did not achieve more than a 0.47 dice score. However, 3D modalities, both standard and residual U-Net have a decent increment in performance without any data augmentation \cite{verma2022automatic}. The shift of modality from 2D to 3D, for Residual U-Net and standard U-Net, had an increment of 7.7\% and 4.3\% of dice score respectively. But, attention U-Net did not succeed in incriminating its performance by shifting from 2D to 3D. The reason behind it might be due to a lack of augmentations and an insufficient number of samples for training\footnote{The authors have experimented with various optimizers, learning schedulers, and many more hyperparameters.}. The results for these models under different performance metrics are illustrated in Table \ref{res-3dunets}.

To specifically study the behaviour of 3D segmentation models we visualized certain test samples one of which is visualized in Figure \ref{fig-3dunetplot}. We have considered two test samples and visualized each plot in three different axes. In each image, the first image describes the Coronal plane and the middle one is the Sagittal plane. Finally, the terminal one is an axial plane. The combination of these three views gives us an estimate of the 3-dimensional pattern of stroke lesions in the brain. As can be seen from Figure \ref{fig-3dunetplot}, U-Net is unable to provide good segmentation outcomes. Though Residual and Attention U-Net were able to segment well but not on par with the mask.

\begin{figure}[!t]
\includegraphics[width=\textwidth]{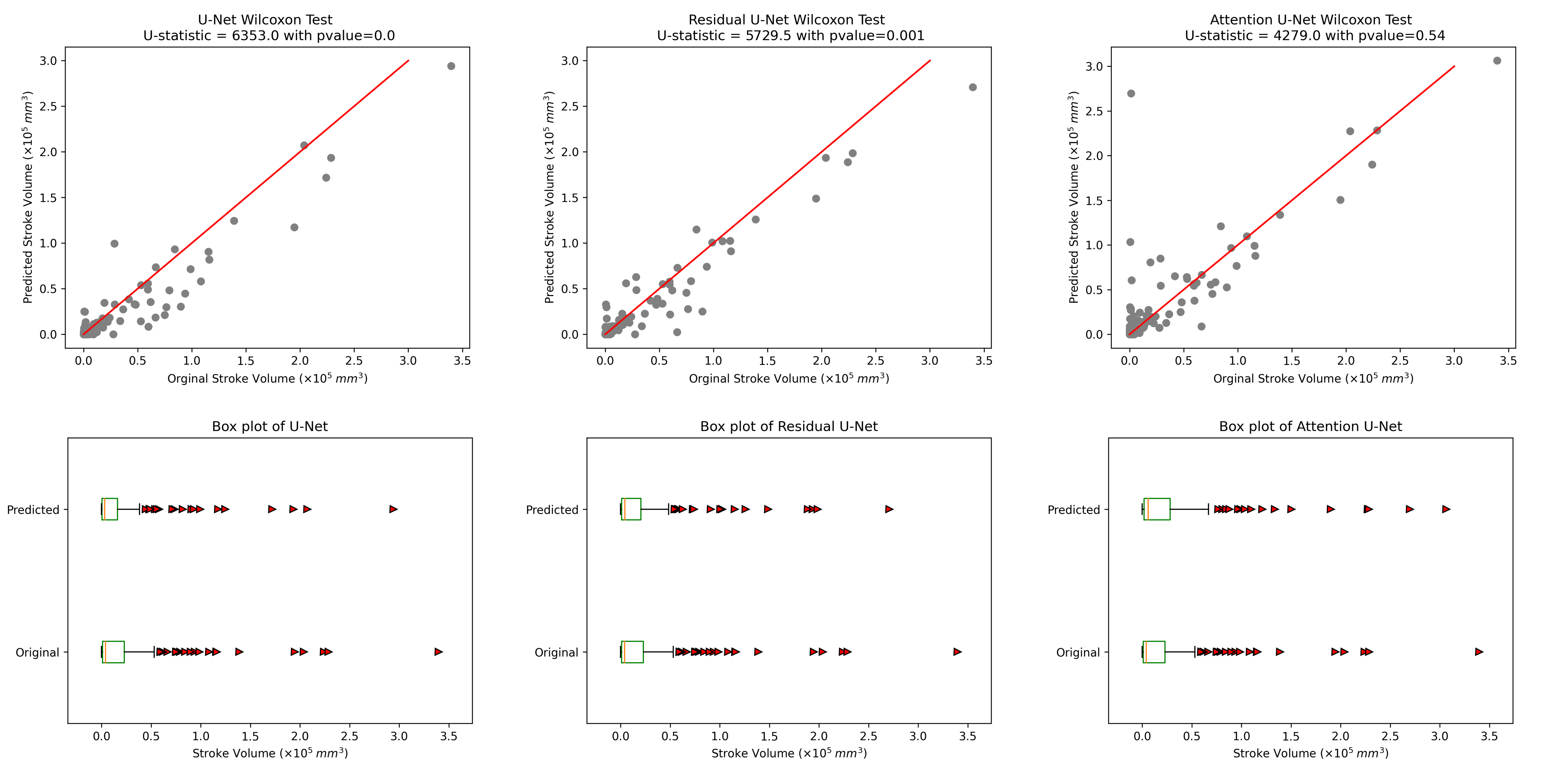}
\caption{The Wilcoxon test is carried out for all three 3D U-Net style architectures. The first row indicates a scatter plot and the second one indicates box plots of the predicted and actual stroke volume of the 3D models respectively. Specifically for the scatter plot. The ideal scenario must be the gray dots aligned with the red line. }\label{wilcox}
\end{figure}

\paragraph{\bf Wilcoxon Signed Rank Test:} Now, we also study the prominence of lesion volume using the ground truths \cite{verma2022automatic}. This is achieved by conducting a statistical test, specifically, the Wilcoxon Signed Rank test, and studying whether the lesion volume distribution patterns for each test subject are similar or not \footnote{This test could be understood as a non-parametric t-test. A detailed premier is provided in the supplementary material.}. Thus, we establish the results for all three 3D U-Nets (Refer Figure \ref{wilcox}). For U-Net and Residual U-Net, the test rejects the null hypothesis and whereas, for attention U-Net it accepts the test with a p-value of 0.54. The detailed results of the Wilcoxon test are detailed in the supplementary material. 
Also, we visualize box plots to assess the original and predicted volume distribution for the test samples as in Figure \ref{wilcox}. 

The distribution of pixels changes after resizing them to a certain shape in the case of 2D. As 2D models are often shrunk and stretched, based on the model, and doing so can misguide the volume calculation. Thus, we cannot estimate the true volumes in such cases and that is the reason we have only reported for 3D models.

\section{Limitations and Future Directions}
In the analysis, we have considered a set of state-of-the-art models. But, there are certain limitations in the current work as described below:
\begin{itemize}
    \item Our motto was to provide standard U-Net style models that are trained on ATLAS v2.0 without any augmentations. Thus, in this current work, we have not used any frameworks such as Deep Medic \cite{kamnitsas2017medic}, nnU-Net \cite{isensee2021nnu}, and MONAI \cite{cardoso2022monai}. 

    \item The current analysis is only done using models that have masks and so we achieved the results with supervision. Later, this work can be extended with weak-supervision \cite{wu2019weakly}, \cite{cao2022ability} or self-supervision \cite{huo2022mapping} approaches which can aid the learning of models in the absence of masks.

    \item This work does not focus on uncertainty or ambiguity in decision-making using generative models \cite{baumgartner2019phiseg} \cite{kohl2018probabilistic} \cite{rahman2023ambiguous}.
    \item Also, we did not address the issue of very small and disconnected lesions \cite{huo2022mapping}. 
\end{itemize}

The limitations described above can be seen as future directions and they could contribute to the progressing field of Neuroimaging for Stroke prediction.

\section{Conclusion}
This paper fundamentally benchmarks variants of U-Net models on the ATLAS v2.0 dataset and deploys standard stroke lesion segmentation models which could be reproducible both for 2D and 3D brain images. We infer that current 2D or 3D brain imaging prediction requires much more attention towards developing hybrid models with the aid of \emph{self-attention} mechanisms to improve the performance of the models. In the future, we tend to develop fine-grained segmentation models with data augmentation, multi-modality (Diffusion Weighted Imaging and T2-FLAIR), and cascaded attention mechanisms. We hope this research could progress and contribute to Stroke prediction.

\section*{Acknowledgement} 
The authors would like to acknowledge Manasa Kondamadugu for her invaluable coordination efforts throughout the project. Additionally, we extend our gratitude to IHub-Data, International
Institute of Information and Technology, Hyderabad for their generous funding and support.

{
\bibliographystyle{splncs04}
\bibliography{samplepaper.bib}

}

\newpage

\section*{Supplementary Material}
\label{app:appendix}

\paragraph{\bf Related Works:}
This section explains each model that was detailed in Table \ref{tab:novelty2d_3d} that elucidates their underlying novelty.

Huo \emph{et al.} \cite{huo2022mapping} presented a stroke lesion segmentation based on the nnU-Net framework and applied it to the ATLAS v2.0 dataset. They also introduced an effective post-processing strategy to enhance segmentation metrics. Their method achieved first place in the 2022 MICCAI ATLAS Challenge with impressive scores, including an average Dice score of 0.6667 and a Lesion-wise F1 score of 0.5643. Verma \emph{et al.} \cite{verma2022automatic} reported for the first time an automated lesion segmentation model on the ATLAS v2.0 dataset which was a 3D U-Net architecture similar to the nnU-Net based framework. The model achieved a Dice similarity coefficient of 0.65. Yu \emph{et al.} \cite{yu2023san} proposed SAN-Net, a self-adaptive normalization network and it also incorporates symmetry-inspired data augmentation (SIDA) for improved generalization for unseen sites. Experimental results show better performance compared to recent methods on the ATLAS v1.2 dataset.

Wu \emph{et al.} \cite{wu2023w} present a novel two-stage network called W-Net for lesion segmentation in ischemic stroke using multi-modal MRI data. W-Net combines CNN-transformer architecture, introduces a boundary deformation module (BDM) to approximate the target boundary, and achieves the best performance in terms of DSC, HD, and F2 metrics, with scores of 61.76\%, 32.47, and 64.60\%, respectively, on the ATLAS v1.2 and ISLES2022 datasets. Du \emph{et al.}  \cite{du2022agmr} proposed AGMR-Net, a novel method for stroke lesion segmentation, addressing intraclass inconsistency and interclass indistinction challenges. AGMR-Net utilizes a coarse-grained patch attention module, a cross-dimensional feature fusion module, and a multiscale deconvolution upsampling module. AGMR-Net achieved impressive results with a high Dice similarity coefficient of 0.594, Hausdorff distance of 27.005 mm, and average symmetry surface distance of 7.137 mm, on the ATLAS v1.2 dataset. 

Sheng \emph{et al.} \cite{sheng2022cross} proposes CADS-UNet, a novel cross-attention and deep supervision UNet, for segmenting chronic stroke lesions from T1-weighted MR images. The model incorporates a cross-spatial attention module to enhance spatial focus, a channel attention mechanism to highlight channel characteristics and deep supervision with mixed loss. The model was evaluated on the ATLAS v1.2 with a Dice Similarity Coefficient (DSC) of 0.5564. Qi \emph{et al.} \cite{qi2019x} introduced X-Net, a depthwise separable convolution-based model with a Feature Similarity Module (FSM). They evaluated X-Net on the ATLAS v1.2, achieving encouraging results with a Dice Similarity Coefficient (DSC) of 0.4867 and an Intersection over Union (IOU) of 0.3723. Zhuo \emph{et al.} \cite{zhou2019d}  introduces D-UNet, a novel architecture combining 2D and 3D convolution in the encoding stage with a new loss function called Enhance Mixing Loss (EML). The proposed method was tested on the ATLAS v1.2 dataset, and it achieved a Dice Similarity Coefficient (DSC) of 0.5349 $\pm$ 0.2763 and a precision of 0.6331 $\pm$ 0.295. Zhang  \emph{et al.} \cite{zhang2020mi} developed a novel stroke lesion segmentation approach called multi-inputs UNet (MI-UNet), which integrates brain parcellation information (GM, WM, LV) with original MR image and achieves notable segmentation performance with a Dice score of 56.72\%, Hausdorff distance of 23.94mm, average symmetric surface distance of 7.00mm, and precision of 65.45\%.

\paragraph{\bf Evaluation Metrics: } For segmenting brain stroke lesions, we employed the mean Dice Similarity Coefficient (DSC), Intersection over Union (IOU), Precision, and Recall as our evaluation metrics. These metrics were utilized to assess the performance of our segmentation models.

\paragraph{Dice Similarity Coefficient (DSC):} For the segmentation problem Dice Similarity Coefficient  (DSC) is one of the important parameters used to assess the similarity between predicted and ground truth pixel or voxels values by measuring the overlap between two sets. Where $X$ is the predicted set of pixels and $Y$ is the ground truth.
$$\text{DSC} = \frac{ 2 (X \cap Y)}{X+Y}$$

\paragraph{Intersection Over Union (IOU):}  IOU is an evaluation metric commonly used in image segmentation to assess the performance of the model.  

$$\text{IOU} = \frac{ (X \cap Y)}{(X \cup Y)}$$

\paragraph{RECALL:} Also known as sensitivity or True Positive Rate $(TPR)$, is an important evaluation metric used for segmentation. It measures the ability of a model to correctly identify positive instances. Recall qualifies how well the model captures the relevant pixel or voxels that belong to the target class. 

$$ \text{Recall} = \frac{TP}{(TP+FN)} $$

\paragraph{PRECISION:} Precision measures how well the model performs in correctly identifying the relevant pixels or voxels that belong to the target class out of all the pixels or voxels predicted as positive.

$$\text{Precision} = \frac{TP}{(TP+FP)}$$

\paragraph{\bf Wilcoxon Signed Rank Test Primer:} 
The Wilcoxon Signed Rank Test is a non-parametric statistical test used to compare two related or paired samples. It assesses whether there is a significant difference between the paired observations. It does this by ranking the absolute differences between paired values, summing positive and negative ranks separately, and comparing these sums using a specific statistical distribution. If the test statistic is significant, it indicates a significant difference between the two groups. It's often used when data doesn't meet the assumptions of parametric tests like the t-test. The null hypothesis of the Wilcoxon test is that there is no difference between the distributions of the two groups, while the alternative hypothesis suggests that there is a significant difference \cite{wilcoxon1970critical}. The Python library SciPy is used to generate the results for the Wilcoxon test: \url{https://docs.scipy.org/doc/scipy/reference/generated/scipy.stats.wilcoxon.html}.

\paragraph{\bf Loss Function}
In medical image analysis, we often use a combination of Dice Loss (DICE) and Binary Cross-Entropy Loss (BCE) for segmentation tasks. Dice Loss quantifies the overlap between predicted and actual regions of interest, while Binary Cross-Entropy Loss measures the dissimilarity between predicted probabilities and ground truth labels. By combining these two losses, we can effectively train deep learning models to segment medical images, helping in stroke lesion segmentation accurately. However, we have provided proportionate weightage for each of the loss functions, and the combined loss is described in the equation (\ref{ourloss}).

\begin{equation}
    \mathcal{L}_{DICE} =  1 - \frac{2 \cdot \sum_{i=1}^{N} p_i \cdot g_i}{\sum_{i=1}^{N} p_i^2 + \sum_{i=1}^{N} g_i^2}
\end{equation}

\begin{equation}
    \mathcal{L}_{BCE} = - \frac{1}{N} \sum_{i=1}^{N} [g_i \cdot \log(p_i) + (1 - g_i) \cdot \log(1 - p_i)]
\end{equation}

\begin{equation}\label{ourloss}
    \mathcal{L}_{OURS} = \gamma \mathcal{L}_{DICE} + (1-\gamma)   \mathcal{L}_{BCE}
\end{equation}

Where \(p_i\) is the predicted probability of a pixel/voxel being in the ROI, \(g_i\) is the corresponding ground truth label (1 for ROI, 0 for background), and \(N\) is the number of pixels/voxels. Here, $\gamma$ is the weightage parameter, and we have obtained the best results for $\gamma =0.9$, and thus, all the models (both 2D and 3D) have used this value as default for a fair evaluation.

\paragraph{\bf 2D Parameter description:}
We implemented five U-Net style architectures for 2D image segmentation, namely, 2D U-Net, ResU-Net, Attention U-Net, TransAttn U-Net, and U-Net Transformer. These models possess approximately 31 M, 32 M, 34 M, 25 M, and 11 M parameters, respectively, and have 4 levels of up and down samplings, except for the U-Net Transformer, which utilizes 3 levels of ups and downsamplings.

A preprocessing step was implemented by cropping the images (coordinates: (10, 40) and (190, 220)) to focus on relevant regions and resizing them to 192x192 pixels via bilinear interpolation, we standardized the input dimensions and emphasized salient features associated with stroke lesions. 
During the training process, we employed the Adam optimizer with an initial learning rate of 0.001. To optimize learning adaptability, we incorporated the "ReduceLROnPlateau" learning rate schedule, which dynamically adjusted the learning. Dropout regularization with a probability of 0.2 was employed to mitigate overfitting during training and these models were trained for 50 epochs, utilizing a batch size of 32 for all models except for the U-Net Transformer, which had a batch size of 16.
The 2D architecture of each model consists of multiple convolutional layers with varying numbers of channels (64, 128, 256, 512, 1024) to effectively capture diverse and intricate features in the data. However, the U-Net Transformer model has (64, 128, 256, and 512) channels. Notably, data augmentations were not applied to the all 2D U-Net models.

\paragraph{\bf 3D Parameter description:}
Here we implemented three state-of-the-art 3D models for image segmentation: U-Net 3D, Res U-Net 3D, and Attention U-Net 3D. Each model's number of learnable parameters was meticulously assessed, revealing U-Net 3D to possess 1.40 M parameters, Res U-Net 3D with 1.42 M parameters, and Attention U-Net 3D featuring 1.61 M parameters. These models were intricately designed with 3 levels of up-and-down samplings.
Regarding input image and mask dimensions, both U-Net 3D and Res U-Net 3D were configured with image dimensions of $144 \times 172 \times 128$, while Attention U-Net 3D adopted the dimensions of $144 \times 176 \times 128$.

The Adam optimizer is employed for training the models, with an initial learning rate of 0.001 to facilitate convergence. The learning rate schedule "Cosine AnnealingLR" is applied to all models, with ADAM optimization. For training and validation batches, a batch size of 4 is utilized for all models. Additionally, the early stopping criterion is set at 100 epochs for all models to control training duration.
Regarding the number of channels used, the models incorporate 16, 32, 64, and 128 channels, allowing them to capture diverse and intricate image features.
To prevent overfitting, a weight decay value of 0.0001 is employed for all models. No data augmentations are applied during training.


\begin{table*}[t!]
\centering
\caption{Wilcoxon and Pearson correlation test for actual and predicted stroke volumes for 3D standard, Residual, and Attention U-Net architectures.}
\label{tab-wilcoxon}
\begin{tabular}{ |c|cc|cc|  } 
\hline
\textbf{Method} & \multicolumn{2}{c|}{\textbf{Wilcoxon Test}} & \multicolumn{2}{c|}{\textbf{Pearson Correlation}} \\ \cline{2-5}
& U-statistic & p-value & statistic & p-value \\ \hline \hline
U-Net \cite{ronneberger2015u} & 6353.0 & $\approx 0.0$ & 0.949 & $\approx 0.0$ \\
Residual U-Net \cite{zhang2018road} & 5729.5 & 0.001 & 0.962 & $\approx 0.0$ \\
Attention U-Net \cite{oktay2018attention} & 4279.0 & 0.540 & 0.844 & $\approx 0.0$ \\   
\hline
\end{tabular}
\end{table*}




\paragraph{\bf Visualization Source:} For, the 3D visualizations the authors have used Ni-Learn framework: \url{https://nilearn.github.io/dev/index.html} and for 2D visualizations Matplotlib framework: \url{https://matplotlib.org/}.

\paragraph{\bf Concerns of HD95 Evaluation Metric:} We are thankful and are obliged from the fruitful reviews provided by anonymous reviewers. One of the reviewers suggested using the HD95 score as one of the evaluation metrics. However, the evaluation metrics used in our study are well-established in the literature and accepted across various domains in medical image segmentation. Apart from these metrics, HD95 provides surface-based information and provides additional insights into the delineation of stroke. Due to time constraints, we were not able to provide them. But, we are very much willing to add those in our future work. 

\vspace{52pt}
\end{document}